\documentclass[floatfix,aps, twocolumn, showpacs, nofootinbib, superscriptaddress]{revtex4} 

\usepackage{graphicx}
\usepackage{epsfig} 
\usepackage{amsmath} 
\usepackage{amsfonts}
\usepackage{amssymb} 
\usepackage{hyperref}
\usepackage{dsfont}

  \newcommand{\beq}{\begin{equation}}
\newcommand{\eeq}{\end{equation}} \newcommand{\bea}{\begin{eqnarray}}
\newcommand{\eea}{\end{eqnarray}}

\begin{document}

\author{Nahuel Freitas} 
\affiliation{Departamento de F\'\i sica, FCEyN, UBA, Ciudad 
Universitaria Pabell\'on 1, 1428 Buenos Aires, Argentina}
\affiliation{IFIBA CONICET, UBA, FCEyN, UBA, Ciudad Universitaria
Pabell\'on 1, 1428 Buenos Aires, Argentina} 

\author{Esteban A. Martinez}
\affiliation{Institut f\"ur Experimentalphysik, Universit\"at Innsbruck,
Technikerstra{\ss}e 25/4, 6020 Innsbruck, Austria} 

\author{Juan Pablo Paz}
\affiliation{Departamento de F\'\i sica, FCEyN, UBA, Ciudad
Universitaria Pabell\'on 1, 1428 Buenos Aires, Argentina}
\affiliation{IFIBA CONICET, UBA, FCEyN, UBA, Ciudad Universitaria
Pabell\'on 1, 1428 Buenos Aires, Argentina} 

\title{Heat transport through ion crystals}


 
\date{\today}

\begin{abstract} 

We study the thermodynamical properties of crystals of trapped ions 
which are laser cooled to two different temperatures
in two separate regions. We show that 
these properties strongly depend on the structure of the 
ion crystal. Such structure can be changed by varying
the trap parameters and 
undergoes a series of phase transitions 
from linear to zig-zag or helicoidal configurations. 
Thus, we show that these systems are ideal candidates to observe and 
control the transition from anomalous to normal heat 
transport. All structures behave as `heat superconductors', with a thermal
conductivity increasing linearly with system size and a vanishing thermal
gradient inside the system. However, zig-zag and helicoidal crystals turn
out to be hyper sensitive to disorder having a linear
temperature profile and a length independent conductivity.
Interestingly, disordered 2D ion crystals are heat insulators. 
Sensitivity to disorder is much smaller in the 1D case. 

\end{abstract}

\maketitle 

\section{Introduction}

Micro- and nano-machines \cite{machines} acting as 
engines and refrigerators \cite{Lutz} are available nowadays. 
These systems may be dominated by  fluctuations as they 
operate far from the thermodynamical limit, in a
regime where classical laws cannot be applied. Thus, a novel
field has emerged, known as quantum thermodynamics
\cite{Mahler, Kosloff}, focusing on the study of the 
emergence of thermodynamical laws as an effective description
obtained from a fundamentally quantum substrate. Quantum 
thermodynamics also focuses on identifying the 
essential resources required to perform certain tasks (extract 
work, conduct energy, etc.) in the regime where the 
otherwise universal laws of thermodynamics cannot be 
directly applied. 

In this article we study energy transport through a quantum system 
that can be controlled to an exquisite degree: a crystal of 
cold trapped ions\cite{Schiffer,Morigi}. Using a new
approach\cite{method,errata}, 
that can be applied to study heat transport 
through general structures, we show that heat flow in ion crystals
depends strongly on the crystal structure, which can be 
experimentally varied. Thus, changing the trapping 
field or the number of ions, the crystal undergoes 
phase transitions that change the nature of the equilibrium
state: The crystal may be a linear (1D) or a zig-zag (2D)
chain, or it can have a helicoidal (3D) structure (we restrict ourselves here to 
study these, the simplest, cases, that have already been 
observed in various laboratories \cite{Morigi,Schiffer}). Trapped ions are promising candidates for 
quantum information processing and crystals with about fifteen ions
have been manipulated to create multipartite entangled states implementing 
small versions of quantum algorithms \cite{Blatt1,Blatt2}. More 
recently, they have been used to simulate frustrated magnetic materials and 
the creation of topological defects during phase transitions
\cite{simul1,simul2}. Their potential to simulate energy flow through complex 
networks has also been noticed \cite{Heffnergroup} and the nature 
of heat conduction have been recently analyzed in the most simple 
(linear) cases\cite{Duan,Briegel,Plenio}). In 
particular, in \cite{Plenio} a toolbox of experimental techniques 
has been introduced to measure not only local temperatures 
but also the heat flow through ion crystals. This would allow phenomena such as anomalous heat conduction (that is, heat conduction not following Fourier's law) to be observed in such systems. In this work we analyze heat conduction in general ion 
crystals using a novel analytical technique. We show that all
crystals display anomalous heat transport. However, a small amount of induced disorder 
can control the transition to normal heat transport in 2D and 3D cystals.
Thus, our results indicate that ion crystals are a promising plataform to study
energy transport in complex structures.
\begin{figure}[h!] \centering
    \includegraphics[width=0.9\columnwidth]{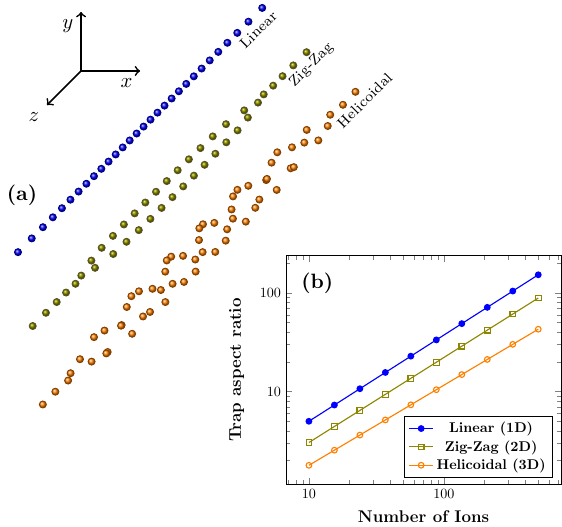} 
        \caption{
        (Color online) (a) Typical crystals with 30, 40 and 60 ions with 1, 2 and 3
        dimensional structures, respectively. 
        (b) Paths in parameter space corresponding to
    structures with the same order parameters (average distance to the
trap axis and mean azimuthal angle between  neighboring ions).  }
\label{fig:structures} 
\end{figure}

In the following section we explain the model employed to describe the
dynamics of the trapped ions and the effects of their interaction with the
reservoirs driving the heat flow. We also describe the method implemented
to calculate the asymptotic state of the crystals.

\section{Model and Methods}
\label{sec:model}

We consider $N$ ions in a Paul trap with harmonic trapping 
potentials both in the axial and transverse directions. The interplay
between Coulomb repulsion and the trapping potentials forms crystals with
variable geometries. For strong transverse confinement, the crystal is
linear (1D). As the transverse potential is relaxed or the number of
ions is increased the crystal undergoes a series of phase transitions.
First, there is a second order phase transition from linear (1D) to a
zig-zag (2D) configuration which is followed by a transition to a
helicoid (3D) and a variety of other shapes. Fully taking  
into account trapping potentials and Coulomb repulsion 
we use an evolutionary algorithm (described in the 
Appendix \ref{sec:phases}) to obtain the equilibrium state of the crystal (in 
a regular desktop computer we can find the equilibrium state of 
crystals with hundreds of ions). In Figure \ref{fig:structures} we 
show three different structures. Zig-zag and
helicoidal structures develop at the center of the crystal and are 
characterized by two order parameters: the mean distance to the axis and
the average azimuthal angle between ions. As shown in the 
Figure \ref{fig:structures}-b, structures with similar order parameters are 
obtained by  appropriately scaling the trap aspect ratio (i.e., 
the ratio between transverse and longitudinal trapping frequencies:
$\alpha =\omega_t / \omega_z$) and the number of ions $N$ (this is explained in
detail in Appendix \ref{sec:phases}).
Once the equilibrium structure is obtained, we quantize the 
oscillations of the ions around equilibrium,
whose dynamics are described by the Hamiltonian 
\begin{equation}
    H_S=\frac{1}{2m}P^TP+\frac{1}{2}X^TVX,
    \label{eq:hamiltonian}
\end{equation}
where the column vector $X=(x_1, \dotsc, x_K)^T$ stores all coordinates
($P$ stores all the momenta; $m$ is the mass of the ions; $K=3N$ is 
the number of degrees of freedom
and the superscript $T$ denotes the transpose). The coupling matrix $V$ 
arises from the second order expansion of the full Hamiltonian. Thus, the 
coupling strengths depend non-trivially on the structure.  

We consider that each transverse coordinate is laser cooled to two different temperatures in the left (L) and right (R)
regions of the crystal, in order 
to induce an energy current through the crystal.
In addition to these engineered reservoirs, other sources of heat could be
considered. Trapped ion strings are subject to heating arising from black body
radiation and RF noise, among other factors. Such mechanisms can be modeled by 
adding an additional thermal reservoir in contact with the whole ion string.
However, for state-of-the-art traps external heating can be reduced to less 
than a couple of phonons per second \cite{labaziewicz2008,niedermayr2014}.
Moreover, most noise sources only contribute to heating of the center-of-mass
motional modes, since typical wavelengths (larger than $1$ cm for frequencies
less than $30$ GHz)
are several orders of magnitude higher than typical ion string lengths 
(not more than $\approx 100$ $\mu$m). Therefore, we will neglect external
heating of the ion string in this work. 
Additionally, off-axis ions will be subject to micromotion coming from the trap
RF fields. The effect of these fields can be incorporated into 
our model as an external periodic drive, which complicates the
analytical solution considerably. However, without going into a
full study of the effect of micromotion on the ion string, we could 
model its effect on heat transport as an additional source of 
heating owning to the factors explained in \cite{wineland1997}.
For the present treatment, we will assume that the heating due to micromotion 
does not affect the transport properties of the crystals.

The evolution of the motional state of a laser cooled ion in a harmonic trap
satisfies a master equation which is analogous to that of a damped harmonic
oscillator coupled to a finite-temperature heat bath \cite{LaserCooling}.
Therefore, to model the cooling process, we couple each of the quantized transverse
coordinates in the left and right regions of the crystal to two bosonic thermal
baths (at temperatures $T_L$ and $T_R$). In order to describe the viscous 
force experienced by the ion, which is proportional to the instantaneous velocity, 
we choose an Ohmic spectral density for each thermal bath. 
Thus, our simplified model consists of a complex network of harmonic degrees of
freedom coupled to two bosonic thermal baths at different temperatures.
This is a generalized version of the usual quantum Brownian
motion (QBM) model which was analitically solved in \cite{PRL}. 

The total Hamiltonian is then $H_T=H_S+H_E+H_{\text{int}}$, where the
environmental Hamiltonian is $H_E=\sum_l H_{E,l}$ with $H_{E,l}=\sum_k
({\pi^{(l)}_k}^2/2m_k+m_k\omega_k^2 {q^{(l)}_k}^2/2)$ and the
interaction is $H_{\text{int}}=\sum_l \sum_{i,k} C_{ik}^{(l)}x_i q^{(l)}_k$ ($q_k^{l}$
and $\pi_k^{l}$ are the coordinate and momentum of the oscillator $k$ of the
environment $l$, respectively, and $C_{ik}^{(l)}$ are coupling constants). The 
asymptotic state of the crystal depends only on two properties of the environments 
\cite{PRL}: The dissipation and the noise kernels, 
defined as  $\gamma(\tau)=\int_0^\infty d\omega
\sum_l I^{(l)}(\omega) \cos(\omega\tau)/\omega$ and $\nu(\tau)=\hbar \int_0^\infty d\omega
\sum_l I^{(l)}(\omega) \coth(\hbar\omega/(2k_BT_l)) \cos(\omega\tau)$. 
The spectral density of each environment is
$I^{(l)}_{ij}(\omega)=\delta_{ij}\delta_{il}\sum_k |C_{ik}^{(l)}|^2
\delta(\omega-\omega_k) /(2m_k \omega_k)$. We assume Ohmic 
environments,i.e.,
\begin{equation}
    I^{(l)}= \frac{2}{\pi}\gamma_0 P_l \frac{\omega \Lambda^2}{\Lambda^2+\omega^2}
    \label{eq:spec_density}
\end{equation}
where $\Lambda$ is a high frequency cutoff, $\gamma_0$ fixes the relaxation 
rate and $P_l$ is the projector onto the coordinates in contact with the $l$-th
environment. These environments
induce a friction force proportional to the velocity.

The interaction with the environment renormalizes the couplings in the
system ($V\to V_R = V - 2\gamma(0)$) and induces friction and diffusion, driving the system to
a Gaussian stationary state. Therefore, a complete description of the
asymptotic state is given by the two point correlation
functions $\langle X X^T \rangle$, $Re \left[ \langle X P^T \rangle \right]$
and $\langle P P^T \rangle$. The heat flow is obtained by
computing the time derivative of the expectation value of the
renormalized Hamiltonian, which is: $d\langle
H_R\rangle/dt = \operatorname{Tr}(V_R \langle X P^T \rangle) / m$. Then the heat 
current entering $L$ (or $R$) is  $\dot Q_{L(R)}= \operatorname{Tr}(P_{L(R)} V_R \langle X P^T
\rangle)/m$, where $P_L$ and $P_R$ are projectors over the transverse
coordinates of the ions in the left or right regions of the crystals. 
This result is intuitive: 
As the force on the $i$-th coordinate is $F_i=-(V_R X)_{i}$, the
expectation value of the power injected in $i$ 
is ${\mathcal P}_i=-\langle (V_R X)_i P_i\rangle/m$. Then, 
because of energy conservation we have 
$\dot Q_i=-{\mathcal P}_i=(V_R \langle X P^T \rangle)_{ii}/m$. 
Momentum correlations define the local kinetic temperature ${\mathcal T}_i$ through the relation 
$\coth(\hbar V_{ii}^{1/2}/(2k_B {\mathcal T}_i))=2 \langle P P^T\rangle_{ii}/(m
\hbar V_{ii}^{1/2})$. In this way, the temperature assigned to a particular
degree of freedom is the one for which the momentum dispersion of a
thermal state of that same degree of freedom matches the momentum dispersion
observed in the asymptotic state of the crystal. In the high temperature
regime, this is simply: ${\mathcal T}_i = \langle PP^T \rangle_{ii} /(mk_B)$.

The covariance matrix of the asymptotic state is well known 
\cite{PRL,Lebowitz}. Denoting $\sigma^{(0,0)}=\langle XX^T\rangle$, 
$\sigma^{(1,1)}=\langle PP^T\rangle$ and $\sigma^{(0,1)}=Re \left[\langle X
P^T\rangle\right]$ 
, we have:
\begin{equation}
\sigma^{(j,k)}= \int_0^\infty d\omega (i)^{k-j} (m\omega)^{j+k} \hat
G(i\omega) \hat \nu (\omega) \hat G(-i\omega).
\label{eq:int_asymp}
\end{equation}
Here, $\hat G(s)$ is the Laplace transform of the Green's function of the system: 
$\hat G(s)=(s^2 m I + V_R+ 2s\hat\gamma(s))^{-1}$, 
and, for an Ohmic environment,  
$\hat \gamma(s)=\gamma_0  P_{T} \Lambda/(s+\Lambda)$ is the Laplace
transform of the dissipation kernel 
($P_{T}=P_L+P_R$). Also, $\hat\nu(\omega)$ is the Fourier transform of the
noise kernel. From Eq. \ref{eq:int_asymp} it is
possible to derive a integral expression for the heat flowing through the
crystal in the stationary state. The following formula is
obtained\cite{PRL}:
\begin{equation}
\begin{split}
    \dot Q = \pi & \int_0^\infty d\omega \; Tr(I_L(\omega) \hat G(i\omega)
    I_R(\omega) \hat G(-i\omega)) \\
    & \times \hbar \omega \left(\coth\left(\frac{\hbar\omega}{k_B T_A}\right) - \coth\left(\frac{\hbar\omega}{k_B T_B}\right)\right).
\end{split}
    \label{eq:int_Q}
\end{equation}

\begin{figure*}
\centering
\includegraphics[width=\textwidth]{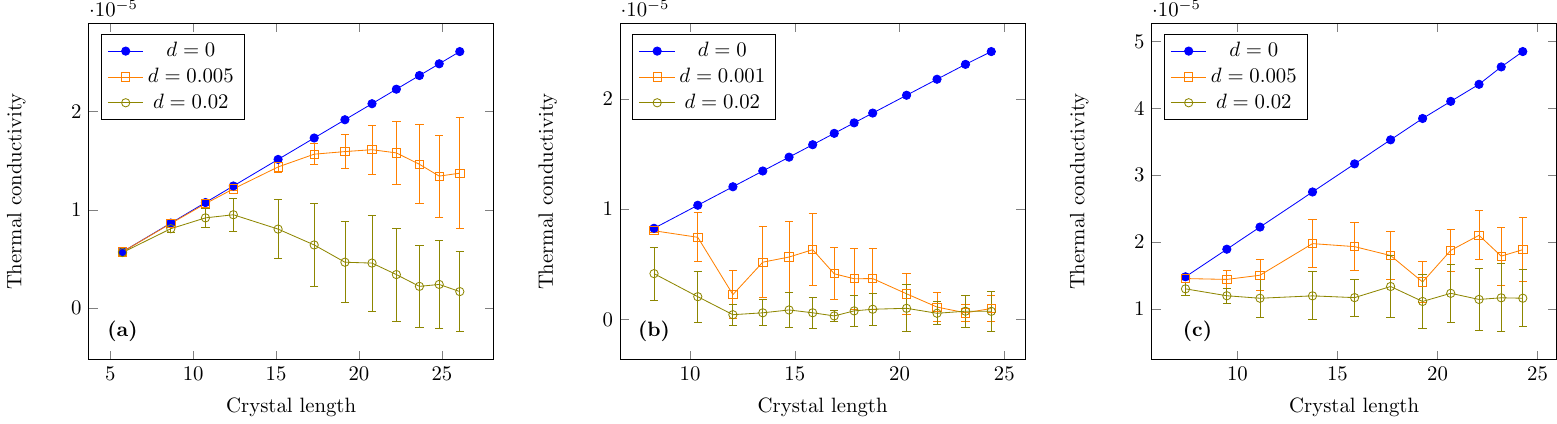} 
\caption{(Color online) Thermal conductivity (in units of $k_B l \omega_z$) as a function of
   crystal length and disorder for 1D lineal (a), 2D zig-zag (b) 
   and 3D helicoidal (c) structures. 
    In all cases we considered between 20 and 200 ions. Points and error bars
correspond to mean values and dispersions over several realizations of
disorder.} 
\label{fig:thermal_cond} 
\end{figure*}

The main ingredient to calculate the two-point correlations and 
the heat current in the stationary state is $\hat G(s)$, 
the Laplace transform of the Green's function. 
To compute the integrals in Eqs. \ref{eq:int_asymp} and \ref{eq:int_Q}
several approximations and techniques are used in the literature\cite{Lebowitz}.
For example, an infinite frequency cut-off is often assumed (which corresponds
to a Markovian approximation, since the dissipation and noise kernels become
local in time). Also, the integrals are usually evaluated numerically, which 
requires the inversion of $-m\omega^2 + V_R + 2i\omega\hat\gamma(i\omega)$ for 
each evaluation point, and therefore those methods are not efficient (nor
accurate) for systems with complex interactions like ion crystals.

Here, we implement a drastically different approach based on  
an analytic formula for $\hat G(s)$, which can be used to 
analytically evaluate the frequency integrals. The method 
is described in detail in \cite{method,errata}. For completeness, we explain here
the main ideas of the method. We consider for simplicity the high-cutoff limit (i.e,
$\Lambda \to \infty$), although the method is also valid 
for an arbitrary cutoff. In that limit $\hat G(s)^{-1} = m s^2  + V_R + 2 s
\gamma_0 P_T $ is a quadratic polynomial in $s$ with matrix coefficients. 
Therefore, to find $\hat G(s)$ it is required to invert a quadratic matrix
polynomial. In analogy with the case of a regular matrix, the inverse
of a quadratic matrix polynomial can be related to the eigenvalues and
eigenvectors of the generalized eigenvalue problem defined by that polynomial. 
Explicitly, it is possible to show that $\hat G(s)$ can be written
as \cite{quadratic}:
\begin{equation}
    \hat G(s) = \sum_{\alpha=1}^{2K} \frac{s_\alpha}{s-s_\alpha} r_\alpha
    r_\alpha^T,
    \label{eq:sum_green}
\end{equation}
where $\{s_\alpha\}$ and $\{r_\alpha\}$ are generalized eigenvalues and
eigenvectors satisfying:
\begin{equation}
    \hat G^{-1}(s_\alpha) r_\alpha = 0,
    \label{eq:eigen}
\end{equation}
which implies $\det(\hat G^{-1}(s_\alpha)) = 0$. Since $\det(\hat G^{-1}(s))$ is a
$2K$ degree polynomial in $s$, there are $2K$ eigenvalues $\{s_\alpha\}$.
Furthermore, since the matrix coefficients of $\hat G^{-1}(s)$ are real, the
eigenvalues and eigenvectors come in complex conjugate pairs. The Laplace
transform of the Green's function $\hat G(s)$ is then expressed in terms of its
poles, which are $\{s_\alpha\}$. In this way the integrals appearing in Eqs.
\ref{eq:int_asymp} and \ref{eq:int_Q} can be evaluated using the residue
theorem. The following result is obtained for the asymptotic covariance matrix:
\begin{equation}
    \sigma^{(j,k)} = 2 \gamma_0 Re \left [ \frac{m^{j+k}}{i^{k-j+1}}
        \sum_{\alpha,\beta=1}^{2K} \omega^{j+k+1}_\alpha
        \omega_\beta
        \frac{r_\alpha^T A
        r_\beta}{\omega_\alpha+\omega_\beta} r_\alpha r_\beta^T
     \right],
    \label{eq:sumCov_high}
\end{equation}
where $A = 2k_B \sum_l T_l P_l$, and
$\omega_\alpha = -i s_\alpha$ are the complex normal frequencies. For the heat
current the result is:
\begin{equation}
    \dot Q = 4\gamma_0^2 \Delta \sum_{\alpha,\beta=1}^{2K} 
    \frac{\omega_\alpha^3 \omega_\beta }
    {\omega_\alpha + \omega_\beta} 
    (r_\alpha^T P_l r_\beta) (r_\beta^T P_{l'}r_\alpha),
    \label{eq:sumQ_high}
\end{equation}
with $\Delta = -2ik_B(T_L-T_R)$. 
Equations \ref{eq:sumCov_high} and \ref{eq:sumQ_high} are valid for high
temperatures. Exact expressions for arbitrary temperature involving the digamma
function can be found in \cite{errata}. Using Eqs. \ref{eq:sumCov_high} and \ref{eq:sumQ_high},
the asymptotic state and heat currents can be evaluated by simply solving a
quadratic eigenvalue problem. In turn, the quadratic eigenvalue problem can be
mapped to a regular (i.e., linear) eigenvalue problem by standard
techniques\cite{quadratic} (at the price of doubling the dimension of the
problem). 

Surprisingly, this method has never been 
used to study transport in harmonic networks. In
simple terms, the method provides the solution 
to a system of $K$ differential equations 
$ m \ddot X+\Gamma \dot X + C X=0$ for arbitrary
non-commuting coupling and damping 
matrices $C$ and $\Gamma$. In the case of spectral densities with finite
frequency cutoff expressions similar to Eqs. \ref{eq:sumCov_high} and
\ref{eq:sumQ_high} can be derived, this time in terms of the eigenvalues and
eigenvectors of a cubic eigenvalue problem\cite{method,errata}.
The method can be used to study transport phenomena in arbitrary harmonic
networks, and it is thus suitable for non-trivial coupling matrices as the ones 
describing ion crystals, that include long-range Coulomb interactions. 

\section{Results}
\label{sec:results}

We now present results for ion crystal with 
up to $N=200$ ions with various structures. We use $m$, the mass of the ions,
as the unit of mass, and $2\pi \omega_z^{-1}$ as the unit of time. The length
unit we use is given by $l=(Q^2/(m \omega_z^2))^{1/3}$ where $Q$ is the electric charge
of the ions (see Appendix \ref{sec:phases}). 

We analyzed the energy flow for the transverse 
motion and considered cases where the environment couples with 
single sites or with extended regions containing 
up to $10$ percent of the crystal (no significant 
differences were found, in accordance with the results in
[\cite{velizhanin2013}] for
the weak coupling regime). We computed the 
thermal conductivity $\kappa$, which is such that
$\dot Q_L=\kappa\Delta  T/L$ where 
$\Delta T=T_R-T_L$ and $L$ is the length 
of the crystal. Fourier's law for macroscopic
heat flow implies that $\kappa$ is $L$ 
independent and also temperature independent. 
We find that fixing $\Delta T$, $\kappa$ 
rapidly becomes independent of the average  
$\bar T$. Also, $\kappa$ becomes independent
of $\Delta T$ for moderately high values of $\bar T$ 
(this behavior is observed for 
temperatures of the order of the frequencies 
in $V_R/m$). Thus, these aspects of Fourier's law 
are valid. All the following results correspond to a regime
in which the heat current is proportional to $\Delta T$, i.e, we 
consider that all the normal modes are thermally excited. 

However, for all structures $\kappa$  
depends linearly on the length, as shown 
in Figure 2. This anomalous behavior is a well known 
property of harmonic chains \cite{Lebowitz,Dhar,Tanos}
that has not yet been experimentally tested. If that
behavior is extrapolated to the thermodynamic limit an infinite thermal
conductivity would be obtained. Therefore, heat could be transported with
the application of a vanishingly small temperature gradient. This fact, and 
the absence of an internal temperature gradient, which is discussed later,
are reminiscent of the behavior of the electric current in superconducting
materials. In \cite{method} it is 
shown that in the weak coupling limit any system which is symmetric with
respect to the interchange of the reservoirs (condition that is fulfilled in our
model) will display a thermal conductivity increasing linearly with the size of
the system. That general scaling law is not longer valid if the 
coupling between system and reservoirs is not small
with respect to the internal couplings in the system (even if the symmetry
condition still holds). 
Since our main interest is to study the effect of disorder on
structures of different dimensionality, we restrict ourselves to the weak coupling regime (we set
$\gamma_0 = 10^{-6}$ for the numerical computations), although the method we use
is valid for arbitrary coupling strength. 

We show that ion crystals are ideal candidates to 
measure and control anomalous transport 
by changing the crystal structure or by adding disorder.
Experimentally, disorder can be implemented in different ways.
One possible approach would be to locally modify the confining potentials.
However, if the potentials are to be tuned using variations in the RF 
confinement voltages, the electrodes have to be small enough to
allow control of individual ions. The electrode size is limited
by the ion-electrode spacing, and the actual state-of-the-art allows
for ion-electrode spacings of $\sim$30-40 $\mu$m [\cite{stick2006ion,seidelin2006}] 
which is larger than typical ion-ion distances of $\sim$10 $\mu$m. 
Improvements in trap miniaturization might render this approach
feasible in the near future. However, disorder can be implemented
with the present technology using site-specific optical dipole forces.
Individual ions in the ion string can be addressed with lasers 
to create an optical lattice with a confinement that can be tuned 
from site to site \cite{enderlein2012single,Plenio}.

In order to study the effects of disorder in a simple way, we numerically 
introduced disorder by modifying the coupling matrix $V$ corresponding to a
particular equillibrium structure. Specifically, we changed 
the pinning potential of $N/2$ randomly selected ions according to the rule 
$V_{ii} \rightarrow (1\pm d)V_{ii}$, where $d$ is a 
measure of disorder. This is expected to be a qualitatively good model of 
particular physical realizations of disorder only for small values of $d$.
However, values of $d$ as small as $0.005$ are enough to control the transition
from anomalous to diffusive heat transport. As shown in Figure 2, linear, zig-zag and helicoidal 
crystals display drastically different behavior as a 
function of disorder. Thus, zig-zag and helicoidal crystals 
are highly sensitive to disorder. In fact, a small $d$ 
turns the zig-zag crystal into a heat
insulator with  $\kappa$ rapidly approaching 
a vanishingly small value. The thermal conductivity of helicoidal crystals approaches a nonzero  
value for long crystals. Hypersensitivity to 
disorder is evident in the dependence of $\kappa$ 
on $d$ for a fixed length $L$. This is shown in
Figure 3-(a) where we see that $\kappa$ rapidly 
decays with  $d$ for 2D and 3D crystals. 
Decay for 1D crystals is clearly much slower.

\begin{figure}[ht!] \centering
    \includegraphics[width=0.95\columnwidth]{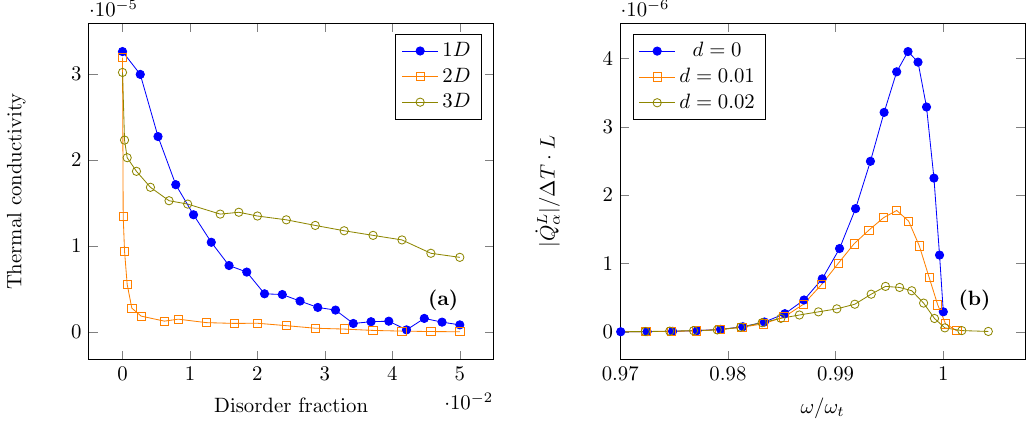} \caption{(Color online) (a) 
Thermal conductivity versus disorder for crystals of 120
ions with different structures. 
(b) Contribution of each normal mode to the thermal 
conductivity in a 1D crystal of 100 ions.
} \label{fig:kappadisorder}
\end{figure}

\begin{figure}[ht!] \centering
\includegraphics[width=0.95\columnwidth]{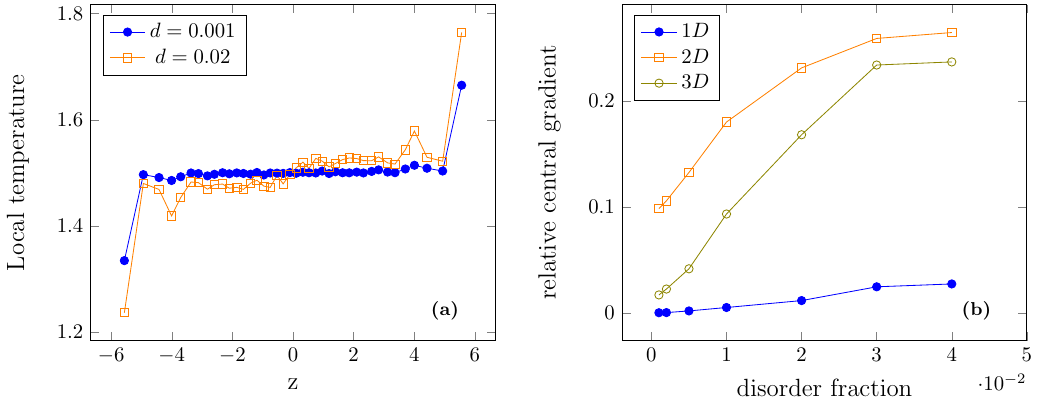} 
\caption{(Color online) (a) Temperature profiles for a helicoidal chain of 40 ions with 
increasing disorder. (b) Central gradient of the temperature 
profile (in units of $\Delta T/L$) as a function of the disorder 
fraction. } 
\label{fig:profile} 
\end{figure}

We also studied the local temperature of the 
transverse motion. As seen in Figure \ref{fig:profile}-a 
the temperature profile strongly deviates from the 
linear behavior predicted by the classical Fourier's law
(we only show the profile for a 3D crystal but no 
substantial differences are seen in 1D or 2D). 
Without disorder the profile is almost planar except 
for the ions in contact with the reservoirs. 
As disorder is introduced, a central temperature gradient 
develops. Thus, deviation from 
Fourier's law can be measured by the central slope 
of the temperature profile, which is shown in Figure
\ref{fig:profile}-b. The central derivative strongly 
depends on the dimensionality: again, 
the zig-zag and helicoidal 
crystals are hiper sensitive to disorder. For small
values of $d$ the central derivative saturates to 
a value which is a significant fraction of the one that correspond 
to a linear interpolation between the temperatures of both reservoirs.

Equations \ref{eq:sumCov_high} and \ref{eq:sumQ_high} enable 
us to estimate the 
contribution of each mode to thermal 
conductivity and temperatures. For example, the contribution of the normal
mode at frequency $\omega = Re(\omega_\alpha)$ to the heat current is:
\begin{equation}
    \dot Q_\alpha = 4\gamma_0^2 \; \Delta \; \omega_\alpha^3 \sum_{\beta=1}^{2K} 
    \frac{ \omega_\beta }
    {\omega_\alpha + \omega_\beta} 
    (r_\alpha^T P_l r_\beta) (r_\beta^T P_{l'}r_\alpha).
    \label{eq:sumQ_mode}
\end{equation}
The behavior of $\dot Q_{\alpha}$ as a function of the mode frequency is 
shown in Figure 3-(b) for different levels of disorder. 
The figure shows that the largest contributions 
come from the modes with higher frequencies. This is expected since 
those are the normal modes with greater amplitude in the ends of the crystals,
and therefore are the ones most coupled to the reservoirs.

\section{Discussion}
\label{sec:discussion}

In summary, we showed that 
ion crystals are excellent candidates to observe 
and control the transition from anomalous to normal 
transport. Thus, by changing the trap parameters 
we induce structural phase transitions which may drive
the crystal to a heat insulating phase (with a zig-zag 
shape), which is a rather remarkable
effect (evidence of the insulating properties of some 
idealized 2D models was presented in \cite{Lebowitz}). 
The toolbox presented in \cite{Plenio} can be used 
not only to measure the heat flow and local temperature 
but also to artificially simulate disorder. In 
this way, the strong dependence of thermodynamical 
quantities on the structure of the crystal 
could be observed with current technologies.
To study this, we implemented a new method 
providing exact formulae for heat currents and 
temperature profiles. All these analytic (exact) 
results depend on generalized eigenvalues and 
eigenvectors of a quadratic problem (which can be easily linearized). 
This work was supported in part by grants from ANPCyT, UBACyT and 
CONICET. EAM is a recipient of a DOC Scholarship from the Austrian Academy of Sciences.

\newpage
\appendix

\section{Structural phases in ion crystals}
\label{sec:phases}

In this appendix we explain how the structural phases corresponding to
1D, 2D and 3D structures were defined. First we present details about
the method used to find the equilibrium configuration of the ion
crystals. We show how structural phase transitions can be determined and
use them to define which are the trap parameters and number of ions
needed to obtain the different structures we used.

\subsection{Equilibrium configuration}

We consider a linear Paul trap with an effective potential that is
harmonic in all directions. The potential energy including Coulomb
repulsion and the effective trap potential for a system of $N$ ions with
mass $m$ and charge $Q$ is:
\begin{equation}
V = \frac{m}{2}\sum_{i=1}^{N} \omega_x^2 x_i^2 + \omega_y^2 y_i^2 +
\omega_z^2 z_i^2 + \frac{1}{2} \sum_{i=1}^{N} \sum_{j\neq i}
\frac{Q^2}{|\bar r_i - \bar r_j|},
\end{equation}
where $\bar r_i =
(x_i,y_i, z_i)$ is the position of the ion $i$ measured from the minimum
of the trap potential. The angular frequencies of the harmonic potential
are $\omega_x$, $\omega_y$ and $\omega_z$. We rewrite the energy in
terms of the parameters $\alpha_x = \omega_x/\omega_z$, $\alpha_y =
\omega_y/\omega_z$ and $q^2 = Q^2/(m \omega_z^2)$:
\begin{equation}
V = \frac{m \omega_z^2}{2} \sum_{i=1}^{N} \left( \alpha_x^2 x_i^2 +
\alpha_y^2 y_i^2 + z_i^2 + \sum_{j\neq i}^N \frac{q^2}{|\bar r_i - \bar
r_j|} \right).
\end{equation}
It is clear from this expression that the
equilibrium configuration of the $N$ ions will only depend on the
parameters $N$, $\alpha_x$, $\alpha_y$ and $q$. We will only consider
the case of a trap with cylindrical symmetry, i.e, $\alpha_x=\alpha_y
=\alpha$. A length scale is fixed by setting $q^2 = 1$.  Thus, the
equilibrium structure is completely determined by $N$ and $\alpha$,
apart from a scaling in the position of all the ions (this scaling can
be performed by varying $\omega_z$ while keeping $\alpha$ to a constant
value).

The determination of $3N$ coordinates $\{(x_i,y_i,z_i)\}$ that
correspond to a global minimum of the energy is a hard problem that even
for small values of $N$ can only be treated numerically. A typical
approach to find a global minimum would be to use some gradient-descent
algorithm combined with some strategy to avoid local minima. We decided
to use a simpler solution based on the differential evolution algorithm
\cite{diffEvol}. This algorithm does not use gradient information,
although it can be taken into account in a very simple way to improve
both convergence and the quality of the solution. This method enable us
to determine equilibrium configurations of crystals with more than 200
ions (600 degrees of freedom) in a modest personal computer.
\begin{figure}[ht!]
\includegraphics[width=0.9\columnwidth]{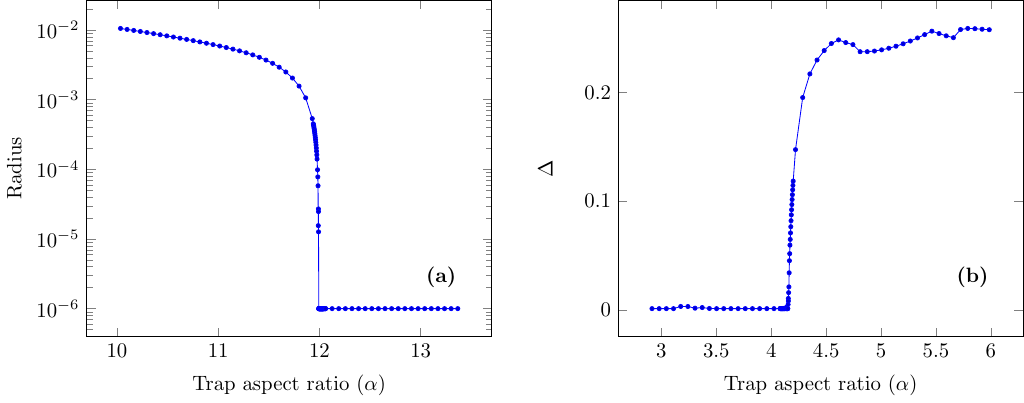}
\caption{\label{fig:phase_trans}(a)Radius of a chain of 30 ions
as the trap aspect ratio is decreased. Transition from 1D to 2D
structures. (b)Minimum longitudinal separations
of the ions ($\Delta$) as the aspect ratio is decreased.} 
\end{figure} 

\begin{figure}
\includegraphics[width=0.9\columnwidth]{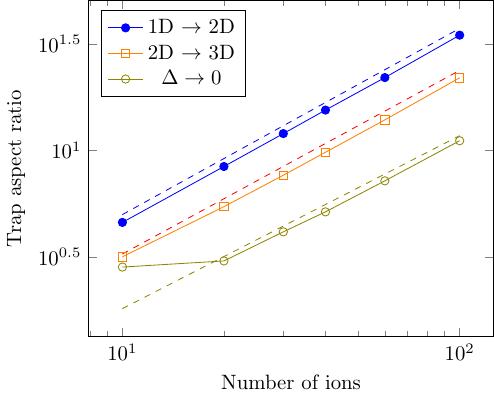}
\caption{\label{fig:phase_trans_points}(Color online) Transition points for several numbers
of ions.} 
\end{figure} 

As is well known, the ion crystals studied here present structural
phase transitions as the trap parameters or the number of ions are
changed \cite{Schiffer}.  We will use these phase transitions to define
`structural phases' in the parameter space spawned by $\alpha$ and $N$.
As an example, Figure \ref{fig:phase_trans}-a shows the
transition from a 1D linear configuration to a 2D zig-zag configuration
in a chain of 30 ions as the transverse trapping potential is relaxed
($\alpha$ is decreased). In this case the order parameter is the chain
radius defined as $R = \max_{1 \leq i \leq N} \{ \sqrt{{x_i^2 +
y_i^2}}\}$. In a similar way it is possible to measure the
transition from 2D configurations to 3D helical configurations
\cite{Schiffer}. Another phase transition occurs if the trap aspect
ratio continues to decrease: beyond some point, the ions in the
equilibrium configuration can no longer be ordered according to their $z$
coordinate, i.e, $z_i \simeq z_j$ for one or more pairs of ions. It is
possible to detect this phase transition by measuring the order
parameter $\Delta = \min_{1 \leq i,j \leq N (i\neq j)}\{|z_i-z_j|\}$, as
shown in Figure \ref{fig:phase_trans}-b.  

The transition points for chains with different numbers of ions are shown
in Figure \ref{fig:phase_trans_points}. As noted in \cite{Schiffer}, the
relation of the critical values of $\alpha$ with the number of ions is
well approximated (for $N >20$) by a simple power law: $\alpha_c \propto
N^\beta$. We estimated the exponent $\beta$ for each
transition. Therefore, we can define power laws $\alpha = c N^\beta$ with the
same exponent $\beta$ but choosing $c$ so that the power laws are
sub-critical. The resulting power laws are shown with dashed lines in
Figure \ref{fig:phase_trans_points}. These paths in parameter space define `structural
phases', in the sense that they determine a family of crystals with
similar structural properties. In table \ref{tab:paths} we give the
values of $c$ and $\beta$ we used to generate 1D, 2D and 3D structures
with different number of ions.

\begin{table}[ht!]
\begin{tabular}{|c|c|c|} \hline &$c$&$\beta$ \\ \hline
1D& 0.67 & 0.873 \\ \hline 2D& 0.44 & 0.861 \\ \hline 3D& 0.28 & 0.811
\\ \hline \end{tabular} \caption{Coefficients of the power laws used to
generate crystal structures of different dimensionality.} 
\label{tab:paths}
\end{table}

\bibliographystyle{unsrt}
\bibliography{references}

\begin{thebibliography}{10}

\bibitem{machines}
G~Cerefolini.
\newblock Nanoscale devices, 2009.

\bibitem{Lutz}
O.~Abah, J.~Rossnagel, G.~Jacob, S.~Deffner, F.~Schmidt-Kaler, K.~Singer, and
  E.~Lutz.
\newblock Single-ion heat engine at maximum power.
\newblock {\em {Phys. Rev. Lett.}}, 109(20):203006, 2012.

\bibitem{Mahler}
J.~Gemmer, M.~Michel, and G.~Mahler.
\newblock Quantum thermodynamics.
\newblock {\em {Lecture Notes in Physics}}, 784, 2009.

\bibitem{Kosloff}
R.~Kosloff.
\newblock Quantum thermodynamics: A dynamical viewpoint.
\newblock {\em Entropy}, 15(6):2100--2128, 2013.

\bibitem{Schiffer}
J.~P. Schiffer.
\newblock Phase transitions in anisotropically confined ionic crystals.
\newblock {\em {Phys. Rev. Lett.}}, 70(6):818, 1993.

\bibitem{Morigi}
G.~Morigi and S.~Fishman.
\newblock Eigenmodes and thermodynamics of a coulomb chain in a harmonic
  potential.
\newblock {\em {Phys. Rev. Lett.}}, 93(17):170602, 2004.

\bibitem{method}
N.~Freitas and J.~P. Paz.
\newblock Analytic solution for heat flow through a general harmonic network.
\newblock {\em {Phys. Rev. E}}, 90(4):042128, 2014.

\bibitem{errata}
N.~Freitas and J.~P. Paz.
\newblock Erratum: Analytic solution for heat flow through a general harmonic
  network [phys. rev. e \textbf{90} , 042128 (2014)].
\newblock {\em Phys. Rev. E}, 90:069903, Dec 2014.

\bibitem{Blatt1}
H.~H{\"a}ffner, W.~H{\"a}nsel, C.~F. Roos, J.~Benhelm, et~al.
\newblock Scalable multiparticle entanglement of trapped ions.
\newblock {\em Nature}, 438(7068):643--646, 2005.

\bibitem{Blatt2}
R.~Blatt and D.~Wineland.
\newblock Entangled states of trapped atomic ions.
\newblock {\em Nature}, 453(7198):1008--1015, 2008.

\bibitem{simul1}
R.~Islam, C.~Senko, W.~C. Campbell, S.~Korenblit, J.~Smith, A.~Lee, E.~E.
  Edwards, C-CJ. Wang, J.~K. Freericks, and C.~Monroe.
\newblock Emergence and frustration of magnetism with variable-range
  interactions in a quantum simulator.
\newblock {\em Science}, 340(6132):583--587, 2013.

\bibitem{simul2}
K.~Pyka, J.~Keller, H.~L. Partner, R.~Nigmatullin, T.~Burgermeister, D.~M.
  Meier, K.~Kuhlmann, A.~Retzker, M.~B. Plenio, W.~H. Zurek, et~al.
\newblock Topological defect formation and spontaneous symmetry breaking in ion
  coulomb crystals.
\newblock {\em {Nature Communications}}, 4, 2013.

\bibitem{Heffnergroup}
T.~Pruttivarasin, M.~Ramm, I.~Talukdar, A.~Kreuter, and H.~H{\"a}ffner.
\newblock Trapped ions in optical lattices for probing oscillator chain models.
\newblock {\em {New J. Phys.}}, 13(7):075012, 2011.

\bibitem{Duan}
G.~D. Lin and L.~M. Duan.
\newblock Equilibration and temperature distribution in a driven ion chain.
\newblock {\em {New J. Phys.}}, 13(7):075015, 2011.

\bibitem{Briegel}
D.~Manzano, M.~Tiersch, A.~Asadian, and H.~J. Briegel.
\newblock Quantum transport efficiency and fourier's law.
\newblock {\em {Phys. Rev. E}}, 86(6):061118, 2012.

\bibitem{Plenio}
A.~Bermudez, M.~Bruderer, and M.~B. Plenio.
\newblock Controlling and measuring quantum transport of heat in trapped-ion
  crystals.
\newblock {\em {Phys. Rev. Lett.}}, 111(4):040601, 2013.

\bibitem{labaziewicz2008}
J.~Labaziewicz, Y.~Ge, P.~Antohi, D.~Leibrandt, K.~R. Brown, and I.~L. Chuang.
\newblock Suppression of heating rates in cryogenic surface-electrode ion
  traps.
\newblock {\em {Phys. Rev. Lett.}}, 100(1):013001, 2008.

\bibitem{niedermayr2014}
M.~Niedermayr, K.~Lakhmanskiy, M.~Kumph, S.~Partel, J.~Edlinger, M.~Brownnutt,
  and R.~Blatt.
\newblock Cryogenic surface ion trap based on intrinsic silicon.
\newblock {\em {New J. Phys.}}, 16(11):113068, 2014.

\bibitem{wineland1997}
D.~J. Wineland, C.~Monroe, W.~M. Itano, D.~Leibfried, B.~E. King, and D.~M.
  Meekhof.
\newblock Experimental issues in coherent quantum-state manipulation of trapped
  atomic ions.
\newblock {\em arXiv preprint quant-ph/9710025}, 1997.

\bibitem{LaserCooling}
J.~I. Cirac, R.~Blatt, P.~Zoller, and W.~D. Phillips.
\newblock Laser cooling of trapped ions in a standing wave.
\newblock {\em {Phys. Rev. A}}, 46(5):2668, 1992.

\bibitem{PRL}
E.~A. Martinez and J.~P. Paz.
\newblock Dynamics and thermodynamics of linear quantum open systems.
\newblock {\em {Phys. Rev. Lett.}}, 110(13):130406, 2013.

\bibitem{Lebowitz}
A.~Chaudhuri, A.~Kundu, D.~Roy, A.~Dhar, J.~L. Lebowitz, and H.~Spohn.
\newblock Heat transport and phonon localization in mass-disordered harmonic
  crystals.
\newblock {\em {Phys. Rev. B}}, 81(6):064301, 2010.

\bibitem{quadratic}
F.~Tisseur and K.~Meerbergen.
\newblock The quadratic eigenvalue problem.
\newblock {\em {SIAM Review}}, 43(2):235--286, 2001.

\bibitem{velizhanin2013}
Kirill~A Velizhanin, Chih-Chun Chien, Yonatan Dubi, and Michael Zwolak.
\newblock Intrinsic thermal conductance, extended reservoir simulations, and
  kramers transition rate theory.
\newblock {\em arXiv preprint arXiv:1312.5422}, 2013.

\bibitem{Dhar}
A.~Dhar.
\newblock Heat transport in low-dimensional systems.
\newblock {\em Advances in Physics}, 57(5):457--537, 2008.

\bibitem{Tanos}
S.~Lepri, R.~Livi, and A.~Politi.
\newblock Thermal conduction in classical low-dimensional lattices.
\newblock {\em {Physics Reports}}, 377(1):1--80, 2003.

\bibitem{stick2006ion}
D.~Stick, W.~K. Hensinger, S.~Olmschenk, M.~J. Madsen, K.~Schwab, and
  C.~Monroe.
\newblock Ion trap in a semiconductor chip.
\newblock {\em {Nature Physics}}, 2(1):36--39, 2006.

\bibitem{seidelin2006}
S.~Seidelin, J.~Chiaverini, R.~Reichle, J.~J. Bollinger, D.~Leibfried,
  J.~Britton, J.~H. Wesenberg, R.~B. Blakestad, R.~J. Epstein, D.~B. Hume,
  et~al.
\newblock Microfabricated surface-electrode ion trap for scalable quantum
  information processing.
\newblock {\em {Phys. Rev. Lett.}}, 96(25):253003, 2006.

\bibitem{enderlein2012single}
M.~Enderlein, T.~Huber, C.~Schneider, and T.~Schaetz.
\newblock Single ions trapped in a one-dimensional optical lattice.
\newblock {\em {Phys. Rev. Lett.}}, 109(23):233004, 2012.

\bibitem{diffEvol}
R.~Storn and K.~Price.
\newblock Differential evolution -- a simple and efficient heuristic for global
  optimization over continuous spaces.
\newblock {\em {Journal of Global Optimization}}, 11(4):341--359, 1997.

\end{thebibliography}

\end{document}